\documentclass{PoS}

\newcommand{\JPsi}{\ensuremath{J\!/\!\psi}}

\title{Quarkonium Production with the CMS Experiment}

\ShortTitle{CMS Quarkonium}

\author{\speaker{Keith A. Ulmer}\\
        University of Colorado\\
        on behalf of the CMS collaboration\\
        E-mail: \email{keith.ulmer@colorado.edu}}

\abstract{Results from studies of quarkonium production are presented from the CMS experiment at the LHC in
proton-proton collisions at $\sqrt{s}$ = 7 TeV. We report measurements of the ratio of 
$\chi_{c2}/\chi_{c1}$ production versus $p_T$ and $\Upsilon(nS)$ production vs rapidity and $p_T$ for the $1S$,
$2S$ and $3S$ states. Reconstruction of $B_c$ mesons is also presented in two decay channels.}

\FullConference{36th International Conference on High Energy Physics,\\
		July 4-11, 2012\\
		Melbourne, Australia}

\begin{document}

\section{Introduction}

The CMS experiment at the LHC pursues a diverse program of heavy quarkonia measurements. 
We present here some recent
highlights of this program, including measurements of the $\chi_{c2}/\chi_{c1}$ production ratio
as a function of $p_T$ and the production of $\Upsilon(1S)$, $\Upsilon(2S)$ and $\Upsilon(3S)$ as
functions of rapidity and $p_T$, including the ratios of the $\Upsilon$ states.
We also present the observation of the $B_c$ meson in two decay channels. 
All measurements were performed 
with the CMS experiment using pp collision data collected at a center-of-mass energy of 7 TeV.

CMS is a general purpose experiment at the Large Hadron Collider\,\cite{CMS}.
The inner detector contains a silicon tracker
composed of pixel layers at radii less than 15\,cm and strip layers out to a radius of
110\,cm.  The central region has 3 layers of pixels and 10 layers of
strips.  The cylindrical geometry of the central
region changes to disks in the $rz$ plane for the forward region.  Each side of the
interaction region contains two endcap pixel layers and up to 12 layers of strips.
The tracker, PbWO$_4$ electromagnetic calorimeter, and
brass-scintillator hadron calorimeter are all immersed in a 3.8~T axial magnetic field.
Muons are measured with detection planes made using three technologies: drift tubes, cathode strip chambers,
and resistive plate chambers.
Muons, electrons and hadrons are tracked within the pseudorapidity region $|\eta| < 2.4$ with
$p_T$ resolution of about 1.5\% for tracks used in the analyses presented here.

\section{$\chi_{c}$ production}

The relative prompt production rate of $\chi_{c2}$ and $\chi_{c1}$ is measured with 4.6 fb$^{-1}$ of data collected at 
LHC at $\sqrt{s}= 7$ TeV~\cite{ChiCPAS}. The two states are reconstructed via the radiative decays 
$ \chi_c \rightarrow J/\psi + \gamma$, with the photon reconstructed through conversions into $e^+e^-$ pairs in 
the tracking detector. The measurements are reported for the kinematic acceptance defined by 
$|y(J/\psi)| < 1.0$ and $p_T(\gamma) > 0.5$~GeV. 
Due to the small mass difference between the $\chi_{c1}$ and $\chi_{c2}$ with the \JPsi, the photons from the 
radiative decays are generally at low momentum, with most between 0.5 and 6.0 GeV in the laboratory frame. Reconstruction
of such low momentum photons with precise energy resolution is only possible through the precise reconstruction of 
electron tracks from photon conversions. 

\begin{figure}[h]
\begin{center}
\includegraphics[clip,width=0.45\linewidth]{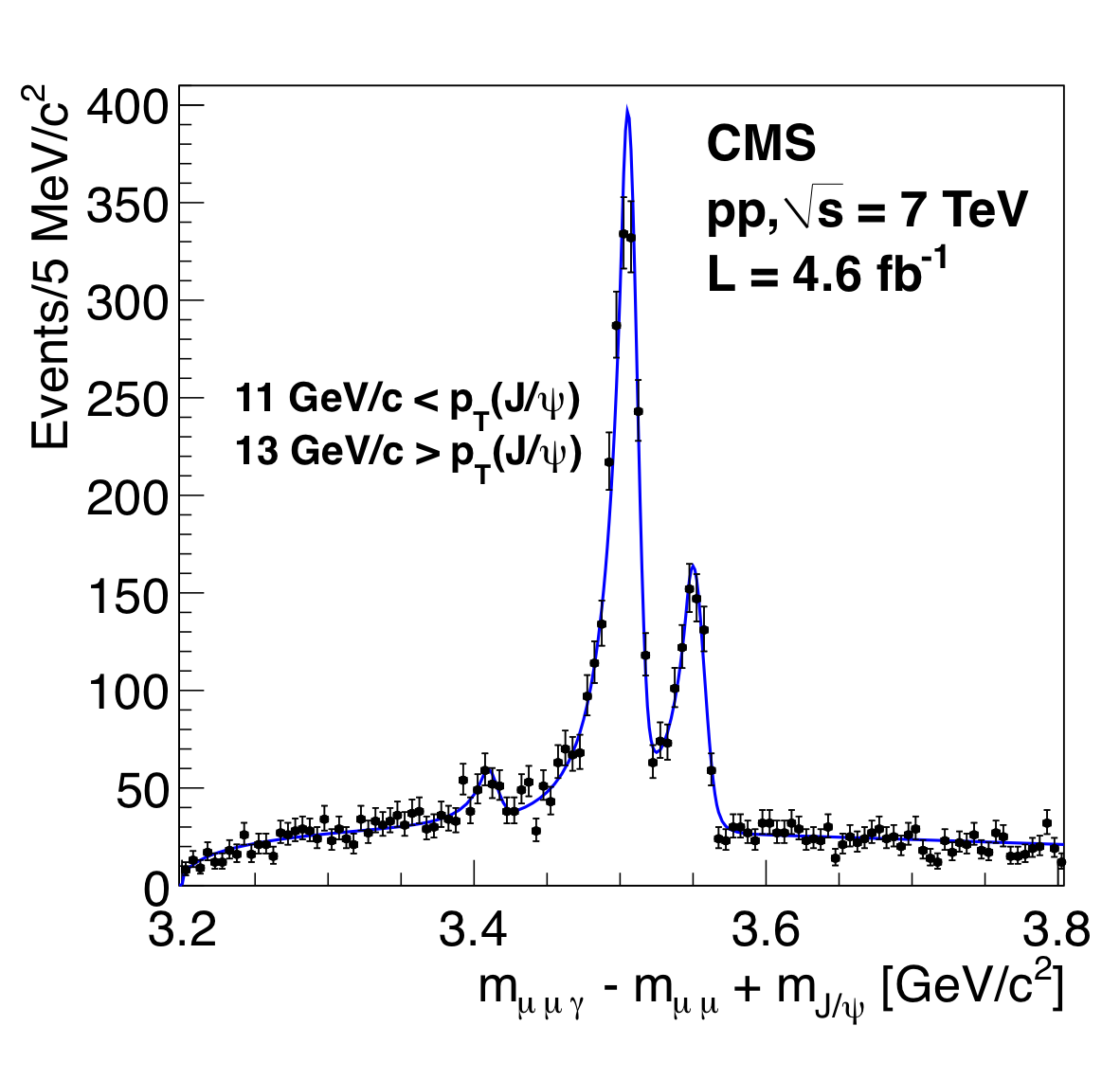}
\includegraphics[clip,width=0.45\linewidth]{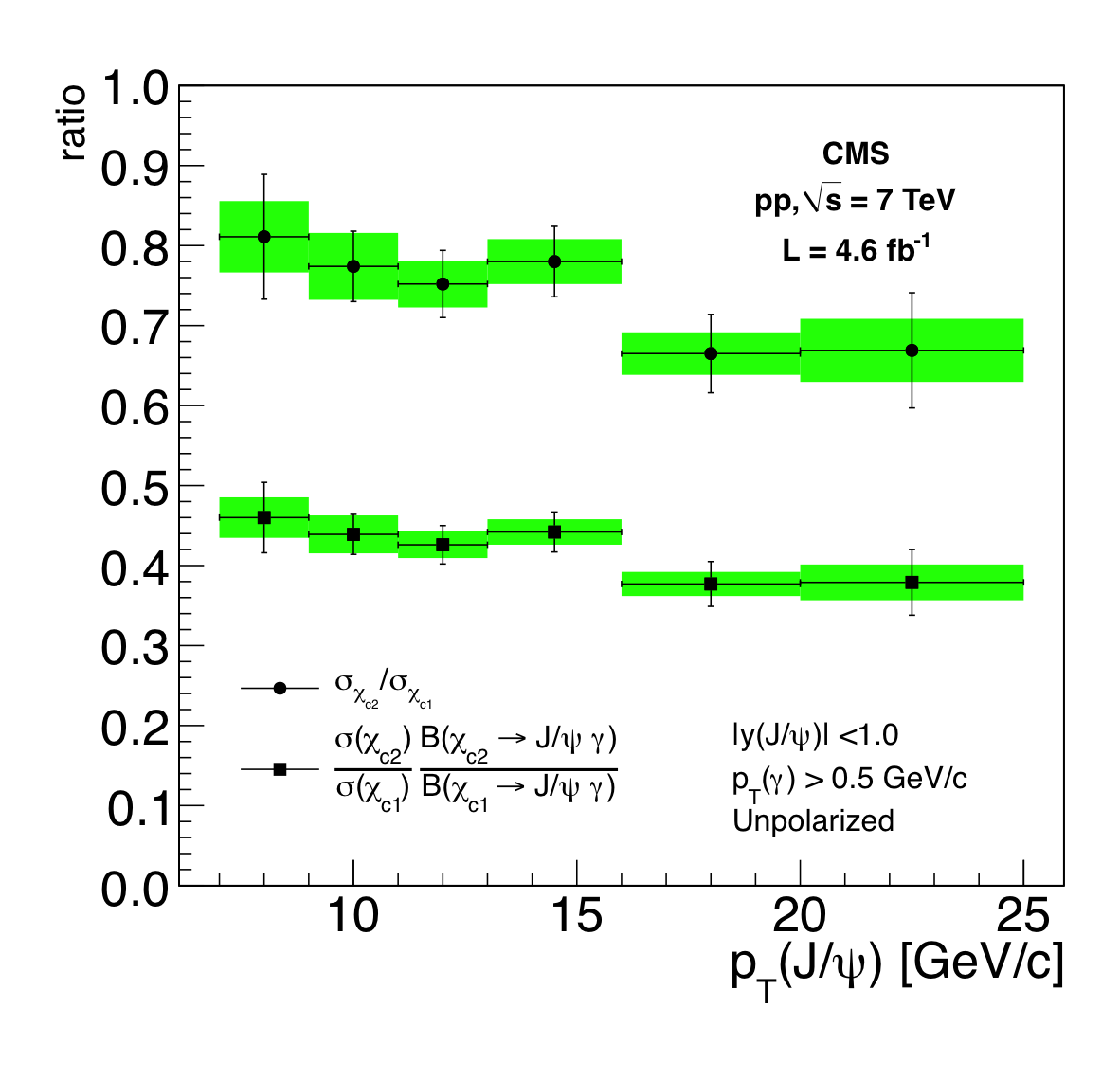}
\caption{Invariant mass distribution for $\chi_c$ candidates with 11 < $p_T(\JPsi)$ < 13 GeV (left) and
$\chi_{c2}/\chi_{c1}$ production results versus $p_T(\JPsi)$ (right).}
\label{fig:ChicResults}
\end{center}
\end{figure}

Oppositely charged muon candidates are fit to a common vertex to produce \JPsi\ candidates, where the two muons 
are also used to trigger the event. To select prompt production only, the proper decay length of 
the \JPsi\ candidate is required
to be less than 30~$\mu$m, which reduces non-prompt contamination to less than $1\%$. Converted photons are
selected by fitting two oppositely charged electron candidates to a common vertex and selecting those that are 
consistent with zero mass and parallel tracks at the point of conversion. The conversion candidate is required to be
consistent with a reconstructed primary vertex in the event. Only the tracker is used in the reconstruction of the
electron candidates, with no information from the calorimeters. Conversion candidates which, when paired with
another photon candidate in the event, are consistent with the $\pi^0$ mass
are rejected. Conversion candidates are then combined with \JPsi\ candidates to form $\chi_c$ candidates. To avoid
large uncertainty in the dimuon mass, the mass difference between the dimuon plus photon system and the dimuon alone
is considered when calculating event yields. Figure~\ref{fig:ChicResults} shows the mass distribution for selected
$\chi_c$ candidates.

\begin{figure}[h]
\begin{center}
\includegraphics[clip,width=0.45\linewidth]{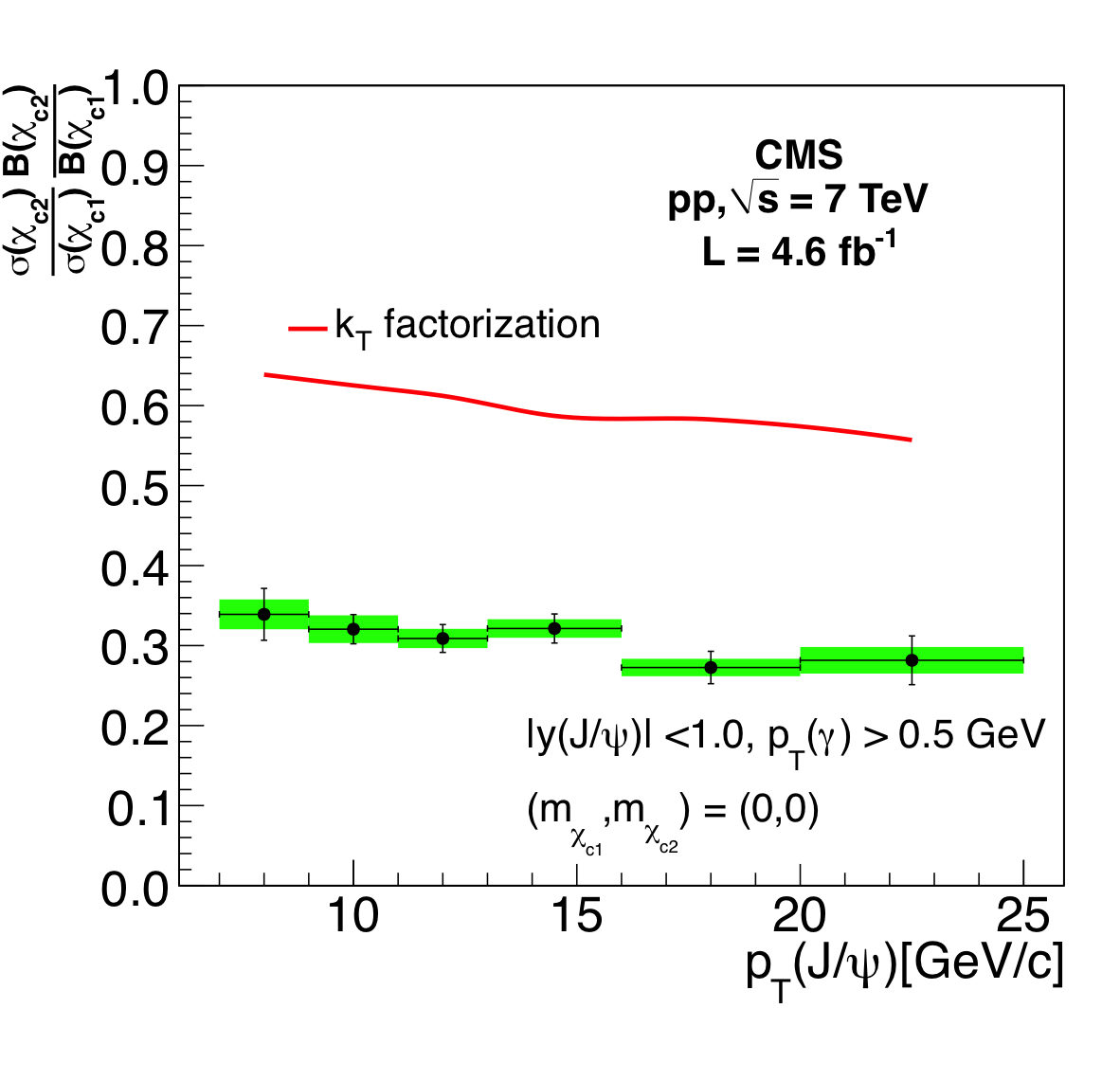}
\includegraphics[clip,width=0.45\linewidth]{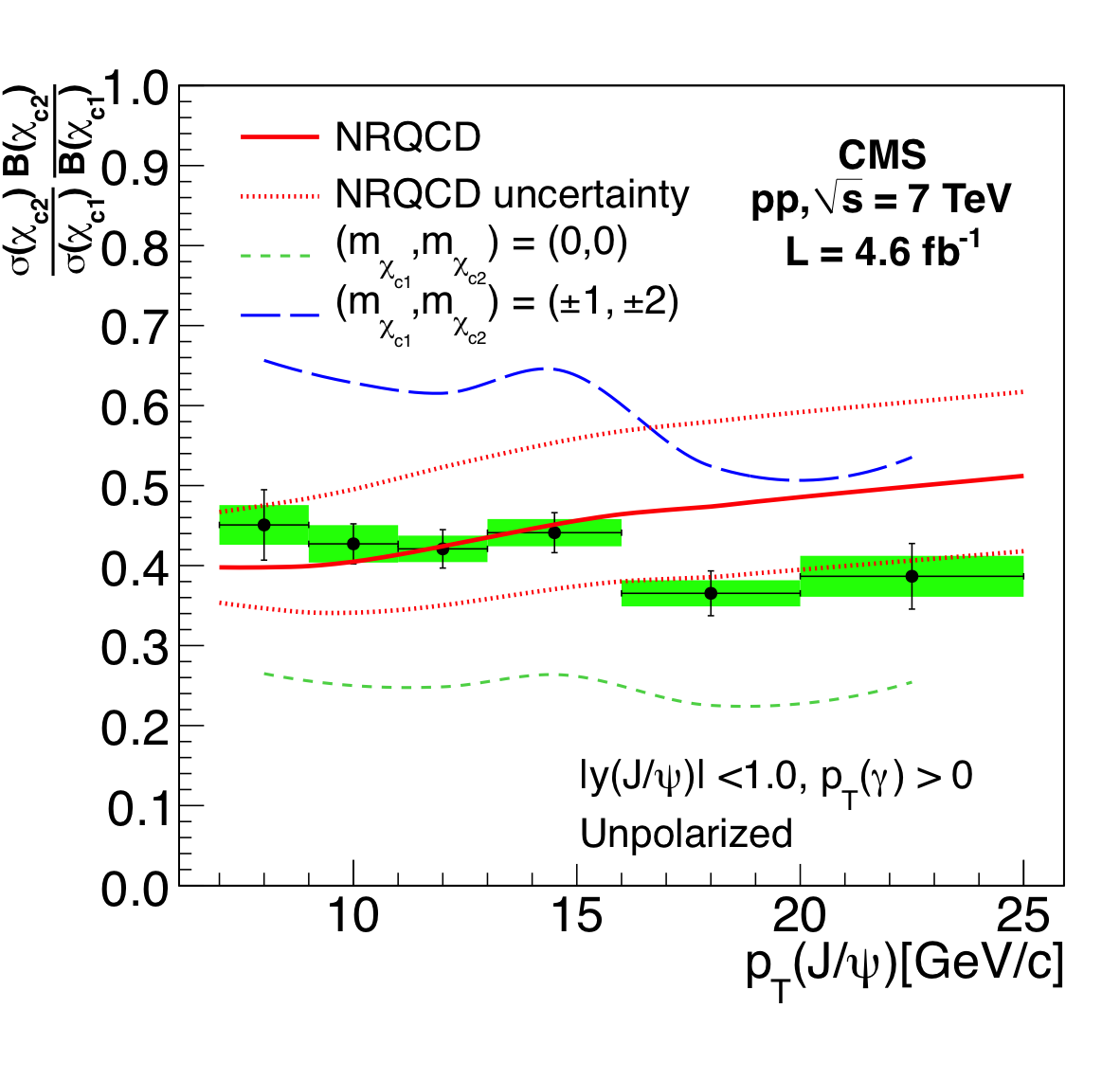}
\caption{Production results for $\chi_{c2}/\chi_{c1}$ versus $p_T(\JPsi)$ compared to various theoretical predictions.}
\label{fig:ChicTheory}
\end{center}
\end{figure}

To measure the ratio of $\chi_{c2}$/$\chi_{c1}$ production, the ratio of efficiencies is required. Many uncertainties
cancel in the use of the ratio of efficiencies. MC simulation is used to determine the ratio of the acceptance and
reconstruction efficiency for the two states, where a maximum difference of $10\%$ is observed. The yields and efficiency
ratios are determined in slices of \JPsi\ $p_T$ to measure the differential production ratio. The results are shown
in Fig.~\ref{fig:ChicResults} with the assumption of unpolarized production of both $\chi_{c1}$ and $\chi_{c2}$. Because
the production polarization of these $\chi_c$ states is unknown, the effects of different polarization assumptions are
reported in~\cite{ChiCPAS} and are found to be as large as $\sim30\%$ on the ratio. Figure~\ref{fig:ChicTheory}
shows the results compared to the predictions from several theoretical models. In order to match the assumptions and
phase space used for the theoretical calculations, the results are reported in several configurations. No single model
is able to completely describe the reported results.

\section{$\Upsilon(nS)$ production}

\begin{figure}[h]
\begin{center}
\includegraphics[clip,width=0.50\linewidth]{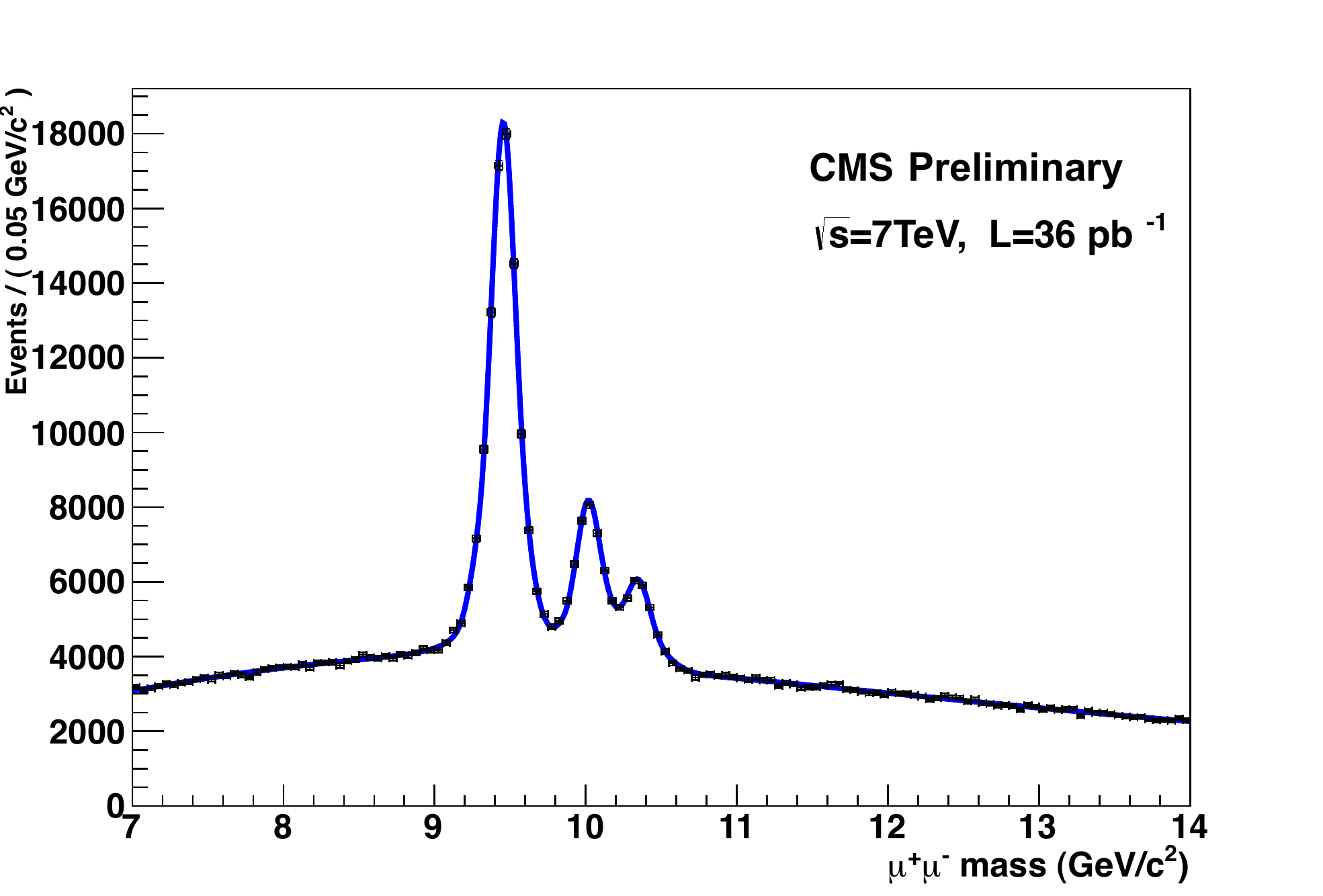}
\includegraphics[clip,width=0.38\linewidth]{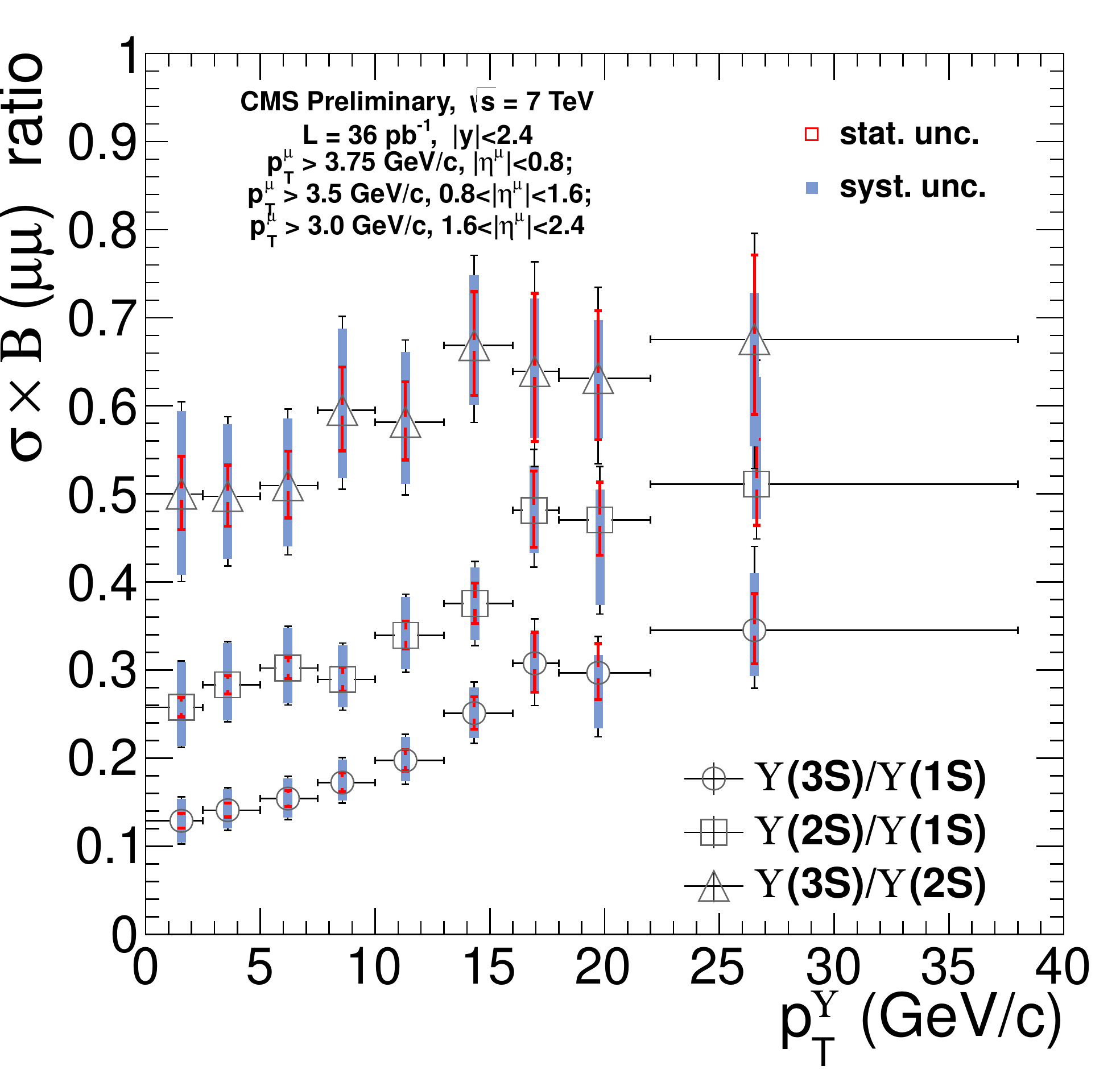}
\caption{Dimuon invariant mass distribution with fitted peaks for the $\Upsilon(1S)$, $\Upsilon(2S)$ and
$\Upsilon(3S)$ states (left) and the $\Upsilon$ production ratios versus $p_T$ (right).}
\label{fig:UpsilonMass}
\end{center}
\end{figure}

The $\Upsilon$ production cross section is measured using $35.8\pm 1.4~{\rm pb}^{-1} $ of proton-proton collisions 
at $\sqrt{s} = 7$~TeV, collected with the CMS detector at the CERN LHC in 2010~\cite{UpsilonPAS}. The  $\Upsilon$
candidates are reconstructed from oppositely charged muon pairs, where the two muons are also used to trigger the
event. A mass resolution of $50\pm2$~MeV is obtained for central muons with $|\eta^{\mu}|<1.0$, while the resolution
is worse for forward candidates. 
Figure~\ref{fig:UpsilonMass} shows the dimuon invariant mass distribution in the region of the $\Upsilon$ resonances.
The efficiency to 
reconstruct an $\Upsilon$ decay is divided into terms for the detector acceptance determined from MC simulation
and the triggering and reconstruction efficiencies determined from \JPsi\ events in data. For the differential cross section
measurement, the efficiencies are determined in bins of $\Upsilon$ rapidity and $p_T$. 

\begin{figure}[h]
\begin{center}
\includegraphics[clip,width=0.45\linewidth]{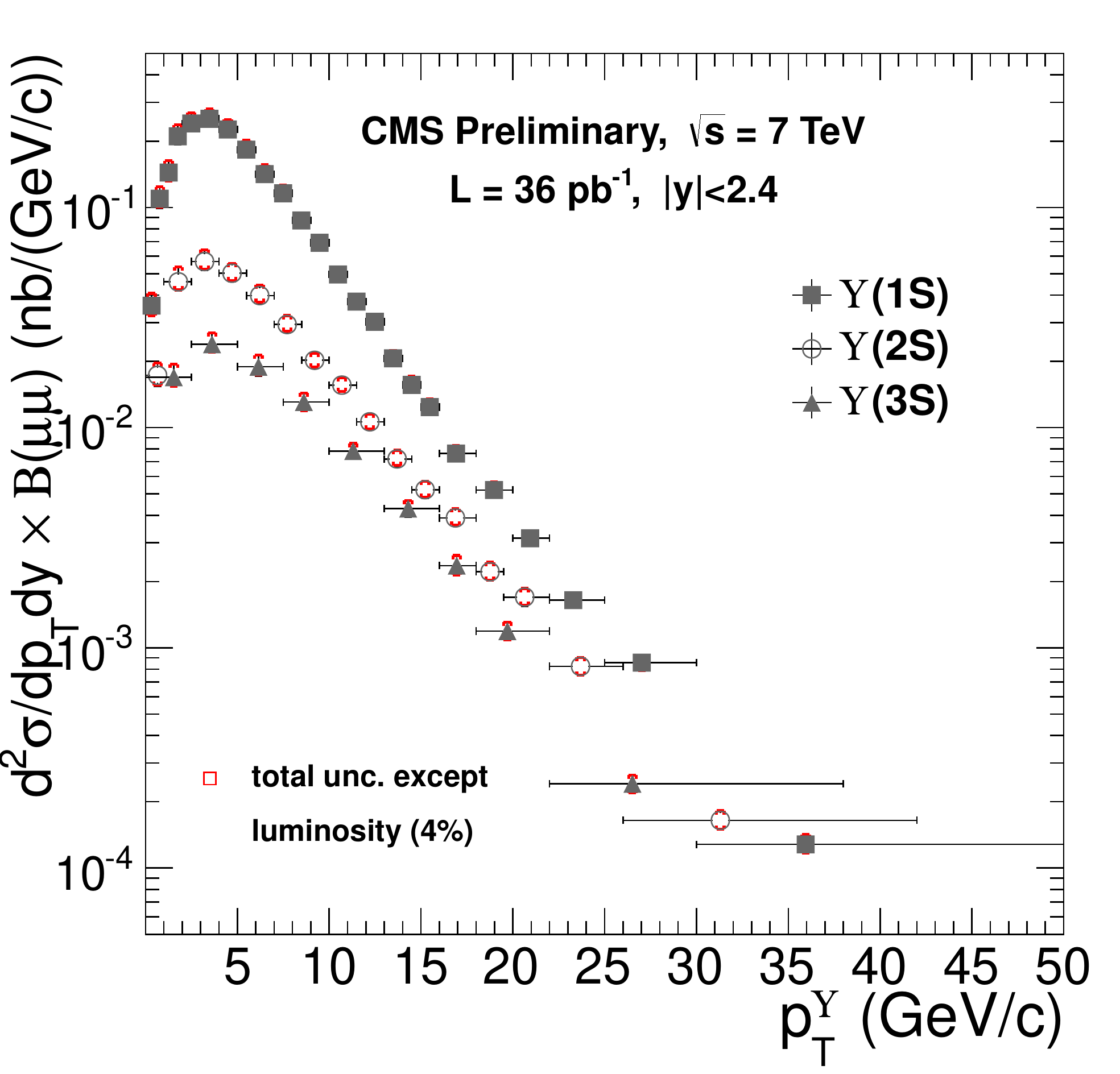}
\includegraphics[clip,width=0.45\linewidth]{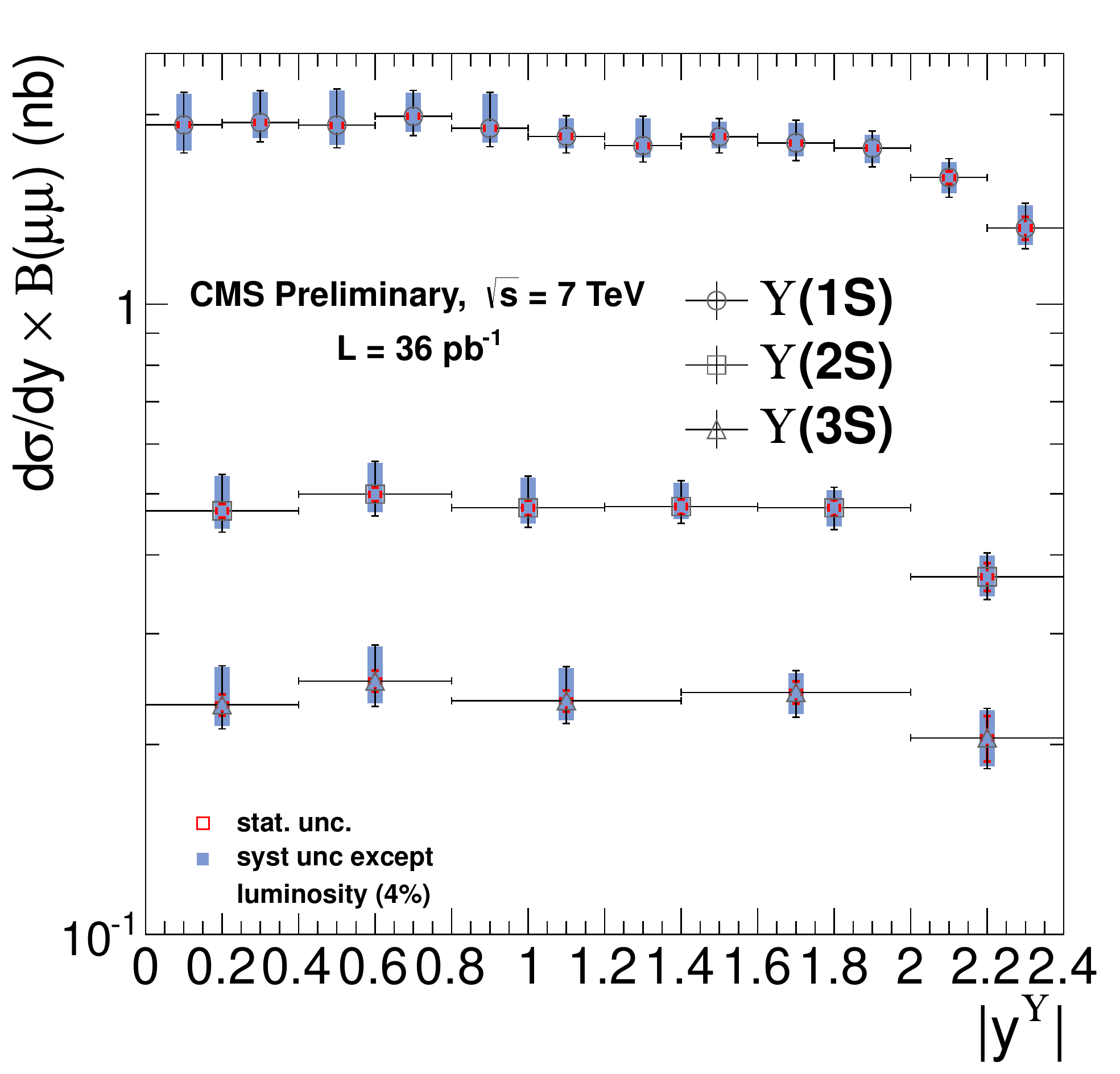}
\caption{Differential production cross section for $\Upsilon(1S)$, $\Upsilon(2S)$ and
$\Upsilon(3S)$ as a function of $p_T$ (left) and rapidity (right).}
\label{fig:UpsilonResults}
\end{center}
\end{figure}

Candidate events are
weighted by the inverse of the efficiency. The weighted distributions are fit in bins of $\Upsilon$ rapidity and $p_T$
to determine the efficiency-corrected yields, which are divided by the luminosity to obtain the differential 
cross sections vs rapidity and $p_T$. Figure~\ref{fig:UpsilonMass} shows the differential cross section ratios between
the $\Upsilon(nS)$ states, while Fig.~\ref{fig:UpsilonResults} shows the differential cross sections versus $p_T$ and
rapidity for the individual states.

The total observed 
yield of the  $\Upsilon$(1S), $\Upsilon$(2S), and $\Upsilon$(3S) reconstructed in the dimuon decay channel are 
$77931\pm 431$, $23847\pm 290$, and $12308\pm 258$, respectively.  
Integrated over the range $p_T$ $<$ 50 GeV and $|y| < 2.4$, assuming unpolarized $\Upsilon$(nS) production, we find
the product of the $\Upsilon$(nS) production cross section and branching fraction to be
\[\sigma(pp \rightarrow \Upsilon {\rm (1S)} X ) \cdot {\cal B} (\Upsilon{\rm (1S)} \rightarrow \mu^+ \mu^-) = (8.56 \pm 0.05^{+0.69}_{-0.56}\pm 0.34)\;\rm{nb}\,,\]
\[\sigma(pp \rightarrow \Upsilon {\rm (2S)} X ) \cdot {\cal B} (\Upsilon{\rm (2S)} \rightarrow \mu^+ \mu^-) = (2.23 \pm 0.03^{+0.19}_{-0.16}\pm 0.09)\;\rm{nb}\,,\]
\[\sigma(pp \rightarrow \Upsilon {\rm (3S)} X ) \cdot {\cal B} (\Upsilon{\rm (3S)} \rightarrow \mu^+ \mu^-) = (1.06 \pm 0.02^{+0.11}_{-0.10}\pm 0.05)\;\rm{nb}\,.\]
where the first  uncertainty is statistical, the second is systematic and the third is associated with the estimation 
of the integrated luminosity of the data sample.
The shapes of the differential cross section measurements versus $p_T$ and $|y|$ are found to be in good agreement
with the predictions from PYTHIA MC simulation.

\section{Observation of $B_c$}

\begin{figure}[h]
\begin{center}
\includegraphics[clip,width=0.45\linewidth]{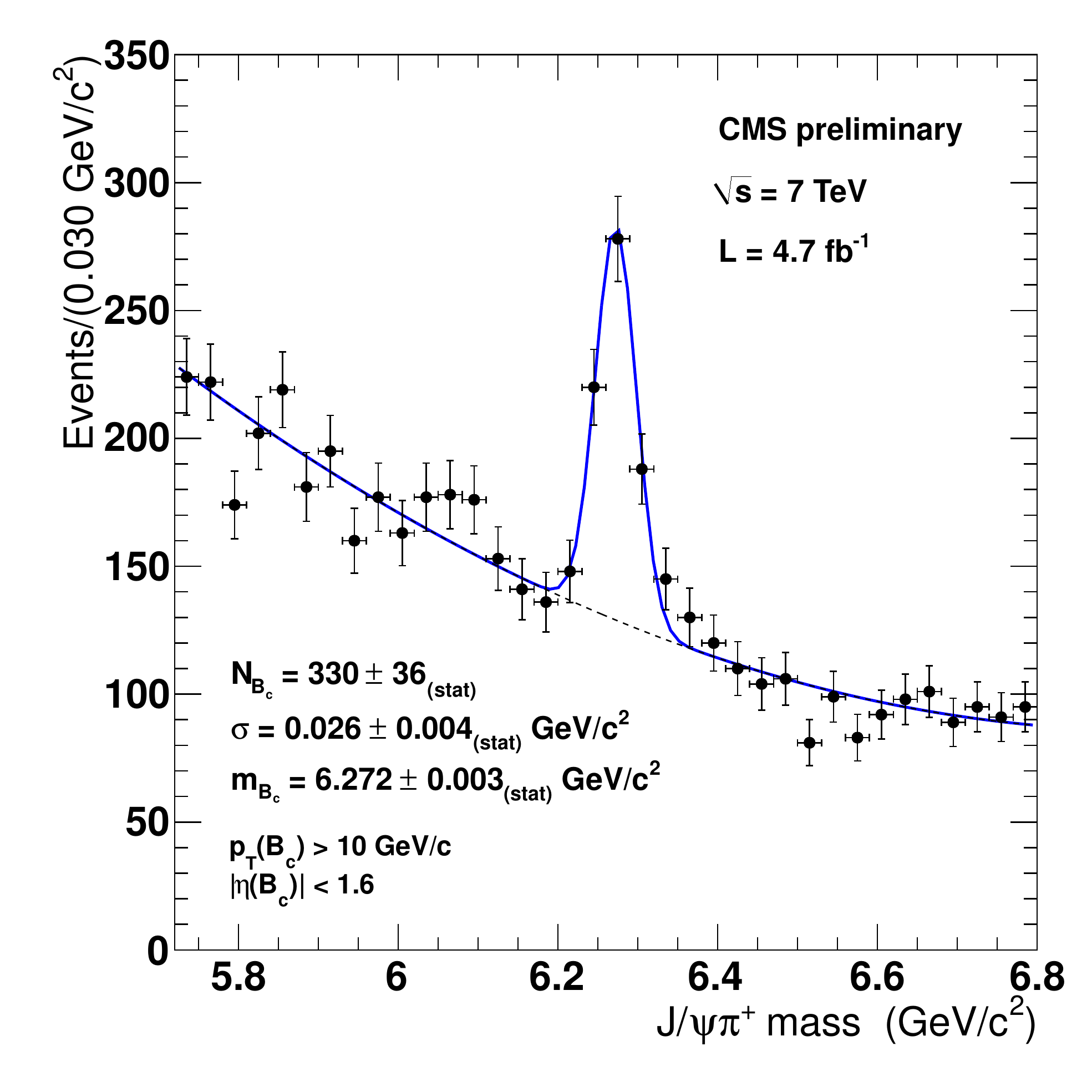}
\includegraphics[clip,width=0.45\linewidth]{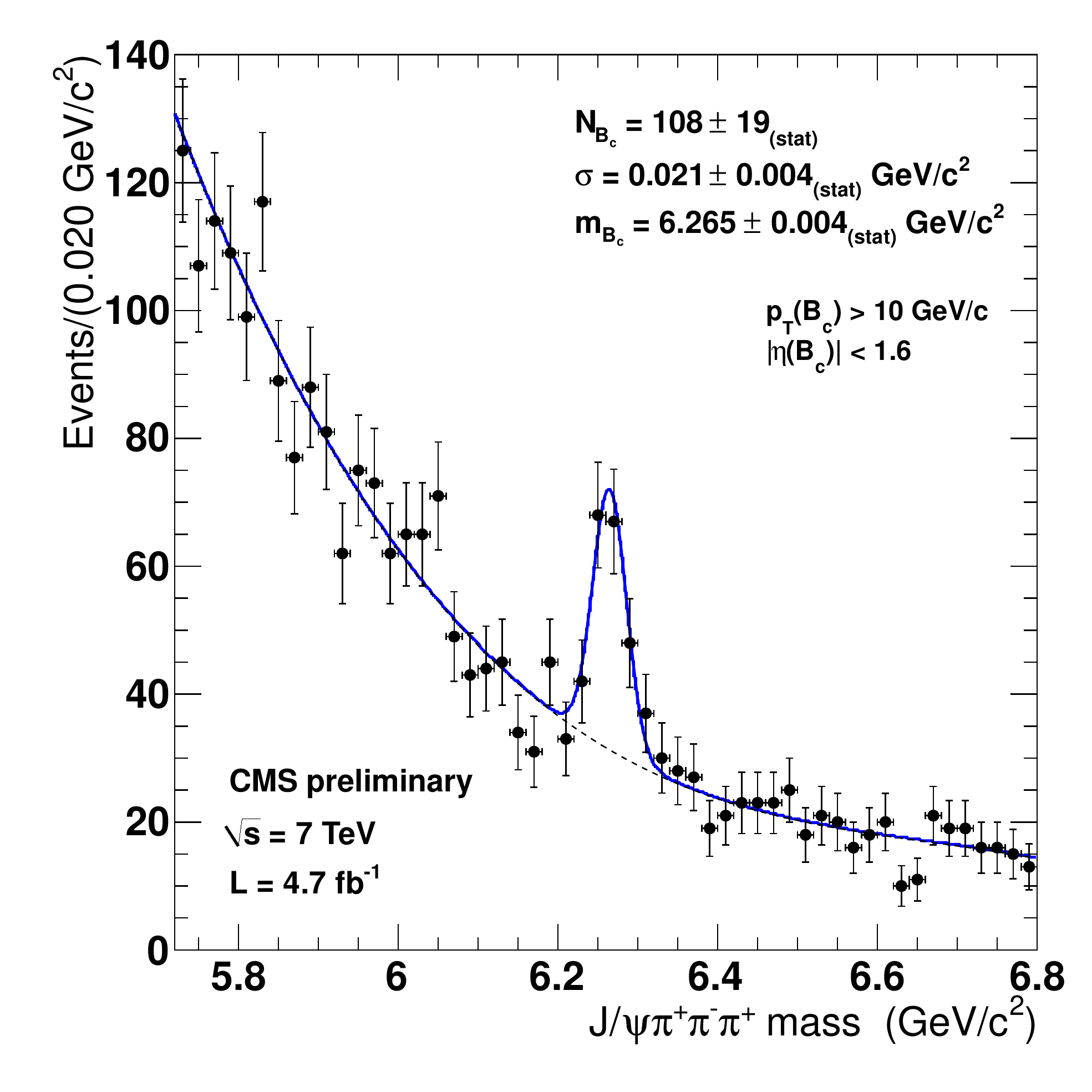}
\caption{Invariant mass distributions for $B_c$ candidates in the $\JPsi\pi^+$
channel (left) and $\JPsi\pi^+\pi^-\pi^+$ channel (right).}
\label{fig:Bc}
\end{center}
\end{figure}

The $B_c^+$ meson, consisting of one beauty quark and one charm quark, presents a unique laboratory
to study heavy quark dynamics. We present the observation of two decay channels at CMS,
$B_c^+\rightarrow\JPsi\pi^+$ and $B_c^+\rightarrow\JPsi\pi^+\pi^-\pi^+$~\cite{BcPAS}. 
In both channels,
\JPsi\ decays to two muons are used to trigger events. Either one or three charged pion candidates
are combined with the \JPsi\ candidate to form $B_c^+$ candidates. A total yield of
330 $\pm$ 36 events are observed for the one pion decay channel, while $108\pm 19$ events are
observed for the three pion decay channel. Figure~\ref{fig:Bc} shows the invariant mass 
distribution for both decay channels.

\section{Conclusion}

A selection of results from the CMS heavy quarkonia physics program is shown, including 
differential measurements of the $\chi_{c2}/\chi_{c1}$ production ratio and the
production of $\Upsilon(1S)$, $\Upsilon(2S)$ and $\Upsilon(3S)$. The observation of 
the $B_c$ meson is also shown. These measurements of heavy quark production further 
our understanding of fundamental quantum chromodynamics, which still does not offer a
complete theoretical picture of the results presented here.

\end{document}